\documentclass[titlepage,11pt]{article}
\usepackage{makeidx,epsfig,amssymb,amsfonts}  
\usepackage{amsmath}
\usepackage{amssymb}
\usepackage{theorem}
\usepackage{enumerate}
\usepackage{url}
\usepackage{color}
\usepackage{lscape}
\usepackage[comma,authoryear]{natbib}
\usepackage{helvet}
\begin{document}

\title{Progressive Mauve: Multiple alignment of genomes with gene flux and rearrangement}

\author{Aaron E.
Darling\thanks{Genome Center and Dept. of Computer Science, University of Wisconsin, Madison, WI
53706 USA }$^,$\thanks{Present address: Institute for Molecular
Bioscience, University of Queensland, St. Lucia QLD 4072, Australia
}$^,$\thanks{To whom correspondence should be addressed: \texttt{darling@cs.wisc.edu}}
\and Bob Mau\thanks{Biotechnology Center and Dept. of Oncology, University of Wisconsin, Madison, WI
53706 USA }
\and Nicole T. Perna\thanks{Genome Center and Dept. of Genetics, University of Wisconsin, Madison, WI
53706 USA }\\
\\
\\
\\
Running title: Multiple genome alignment
lateral gene transfer, homologous recombination, population genomics}

\maketitle              

\begin{abstract}

Multiple genome alignment remains a challenging problem.  Effects of recombination including rearrangement, segmental duplication, gain, and loss can create a mosaic pattern of orthology even among closely related organisms.  We describe a method to align two or more genomes that have undergone large-scale recombination, particularly genomes
that have undergone substantial amounts of gene gain and loss (gene flux).  The method
utilizes a novel alignment objective score, referred to as a sum-of-pairs breakpoint score.  We also apply a probabilistic alignment filtering method to remove erroneous alignments of unrelated sequences, which are commonly observed in other genome alignment methods.

We describe new metrics for quantifying genome alignment accuracy which
measure the quality of rearrangement breakpoint predictions and indel predictions.
The progressive genome alignment algorithm demonstrates vastly improved accuracy
over previous approaches in situations where genomes have undergone realistic amounts
of genome rearrangement, gene gain, loss, and duplication.

We apply the progressive genome alignment algorithm to a set of 23 completely sequenced
genomes from the genera \textit{Escherichia}, \textit{Shigella}, and \textit{Salmonella}.
Analysis of whole-genome multiple alignments allows us to extend
the previously defined concepts of core- and pan-genomes to include not only annotated genes, but also
non-coding regions with potential regulatory roles.
The 23 enterobacteria have an estimated core-genome of 2.46Mbp conserved among all taxa and a pan-genome
of 15.2Mbp.  We document substantial population-level variability among these organisms
driven by homologous recombination, gene gain, and gene loss.  Interestingly, much
variability lies in intergenic regions, suggesting that the Enterobacteriacae may exhibit
regulatory divergence.  In summary, the multiple genome alignments generated by our
software provide a platform for comparative genomic and population genomic studies.

Free, open-source software implementing the described genome alignment approach is available from
\url{http://gel.ahabs.wisc.edu/mauve}.

\end{abstract}

\section*{INTRODUCTION}
Multiple genome alignment is among the most basic tools in the
comparative genomics toolbox, however its application has been
hampered by concerns of accuracy and practicality~\citep{Kumar2007,Prakash2007,Lunter2007}.
Accurate genome alignment represents a necessary prerequisite for myriad
comparative genomic analyses.

The definition of a ``genome alignment'' has been the source of some confusion in
comparative genomics.  Genome alignment goes beyond basic identification of
homologous subsequences to subclassify homology relationships by specific types of
evolutionary history~\citep{Dewey2006}.
At a gross level, subtypes of homology include orthology, paralogy, and xenology~\citep{Fitch2000}.
Paralogs can be further classified as in- and out-paralogs, while orthologs can be
further distinguished as single-copy orthologs (1-to-1 orthologs), tandemly repeated orthologs (co-orthologs), positional,
and general orthologs.  For the present work, we are interested in genome alignments that distinguish
single-copy orthologous and xenologous sites from the remainder of homologous and unrelated sites.
Downstream application of recombination detection methods can distinguish single-copy orthologs from xenologs.

Early genomic studies on \textit{E. coli} revealed substantial gene content variability
among individual \textit{E. coli} isolates~\citep{Perna2001,Welch2002}.  Since then, gene content variability has been
reported as a common feature in numerous other microbial species~\citep{Tettelin2005,Hogg2007,Hsiao2005,Vernikos2007}.
It appears that microbial populations undergo vast amounts of gene flux and homologous recombination~\citep{Mau2006},
although systematic studies have been limited to gene-based methods by the difficulty of complete and accurate multiple genome alignment.

Approaches to whole-genome alignment typically reduce the
alignment search space using anchoring heuristics~\citep{Ogurtsov2002,lagan,mga,mavid,tba} or banded dynamic programming~\citep{Chao1995}.  Anchoring heuristics appear to provide a good tradeoff between speed and sensitivity.
Most anchored alignment methods assume that the input sequences are free from genomic rearrangement.
As such, a separate synteny mapping algorithm must be applied to map collinear homologous segments among
two or more genomes prior to alignment.  Synteny mapping approaches are too numerous
to list, however most involve computing
reciprocal best BLAST hits on putative ORFs, with BLAST hits filtered by e-value thresholds,
coverage thresholds, and uniqueness criteria.  Some synteny mapping methods apply genomic context to help resolve ambiguous orthology/paralogy relationships, and others use probabilistic transitive homology approaches to infer orthologs among distantly related taxa~\citep{OrthoMCL}.

Integrated approaches to synteny mapping and alignment have been proposed, most of which operate on pairs of genomes~\citep{mummer3,slagan,Vinh2006,Swidan2006}.  Research into multiple alignment with rearrangements has been limited, although some progress has been made~\citep{mauve,Treangen2006,aba,mulan,Phuong2006}.
Apart from greater ease-of-use, integrated synteny mapping and alignment methods have the advantage of providing more
accurate inference because the alignment can influence the synteny map and vice-versa.

In the present work, we introduce a novel computational method to construct a multiple
alignment of a large number of microbial genomes which have undergone gene flux
and rearrangement.  Unlike our previous alignment method~\citep{mauve}, the new method aligns regions conserved among subsets of taxa.  Three algorithmic innovations factor strongly in our method's ability to align genomes with variable gene content and rearrangement.  The first is a novel objective function, called a sum-of-pairs breakpoint score, to score possible configurations of alignment anchors across multiple genomes.  Our second algorithmic contribution is a greedy heuristic to optimize a set of anchors under the sum-of-pairs breakpoint score.  We demonstrate that most anchored alignment techniques suffer a bias leading to erroneous alignment of unrelated sequence in regions containing differential gene content. Our final algorithmic contribution is application of a homology hidden Markov model (HMM) to reject such erroneous alignments of unrelated sequence.

We compare accuracy of existing methods and the new alignment method on datasets simulated to encompass a broad range of genomic mutation types and rates, including inversion, gene gain, loss, and duplication.  We then apply the multiple genome alignment method to a group of 23
finished genomes in the family Enterobacteriacae (Supplementary Table~\ref{table:supp_enterics}).  We precisely identify the core- and pan-
genomes of this group, and report basic analysis of gene flux patterns in Enterobacteriacae.

\section*{METHODS}
\vspace{-0.2cm}
An overview of our method as applied to three hypothetical genomes appears in Figure~\ref{fig:algorithm} and is presently described in detail.

\subsubsection*{Notation and assumptions}
\vspace{-0.2cm}
Our genome alignment algorithm takes as input a set of $G$ genome
sequences $g_1, g_2, \dots g_G \in \mathbf{G}$.  We denote the length of
genome $i$ as $|g_i|$.  Contigs in
unfinished or multi-chromosome genomes are concatenated to form a single coordinate system.
Our method computes alignments along a rooted guide
tree $\Psi$ which is computed from $\mathbf{G}$, and we use $\Psi_n$ to denote an arbitrary node in $\Psi$.
As $\Psi$ is a rooted bifurcating tree, an internal node $\Psi_n$ always
has two children, which we refer to as $\Psi_n.c_1$ and $\Psi_n.c_2$.
Furthermore, we
define the set of leaf nodes at or below $\Psi_n$ as $\mathcal{L}(\Psi_n)$.
Each leaf node represents a genome from
the set of input genomes $\mathbf{G}$.  We use set theoretic notation in some places, where $\smallsetminus$ indicates removal of members from a set.

Various default parameter settings in our software implementation
depend on the average length of input genome sequences.  We refer to
the average genome length as $AvgSize(\mathbf{G})$.

\subsubsection*{Local multiple alignments as potential anchors}
\vspace{-0.2cm}
We identify local multiple alignments (LMA) as potential anchors using
families of palindromic spaced seed patterns~\citep{ref-procrast} in a seed-and-extend
hashing method (see Appendix of~\citet{mauve}).  A spaced seed pattern~\citep{patternhunter}
identifies the location of $k$-mers in the input genomes that have identical nucleotide sequence
except that a small number of mutations are allowed at fixed positions.  For example, the seed pattern 11*11*11 would
identify matching oligomers of length 8 where the 3rd and 6th positions are degenerate.  The number of 1's in the
seed pattern is commonly referred to as the \textit{weight} of the seed pattern, and a pattern is said to be \textit{palindromic} if the pattern is identical when read forward or in reverse.  A seed family is a collection of seed patterns that when used in conjunction provide improved matching sensitivity, and such families have been
previously demonstrated to~offer excellent speed and sensitivity~\citep{Kucherov2005}.

To minimize compute time and focus anchoring coverage on single-copy regions,
our method only extends seeds which are unique in two or more genomes.  By default, we use
seed patterns with weight equal to
$\log_2 ( AvgSize(\mathbf{G}) / 1.5 )$.  The resulting local multiple alignments are ungapped
and always align a contiguous subsequence of two or more genomes in $\mathbf{G}$.  Any given local multiple alignment
$m$ can be described formally by its length $m.Length$ and a tuple $\langle x_1, x_2, \dots, x_G \rangle$ where
$x_i$ is an integer denoting the left-end coordinate of the LMA in genome $g_i$.  Genomic coordinates are assumed to start at 1, with negative values denoting alignments to the reverse strand,
and when $x_i$ takes on a value of 0, it indicates that genome $g_i$ is not part of the LMA.~
The LMAs found by our procedure are ungapped alignments of unique subsequences and thus are similar to multi-Maximal-Unique-Matches
(multi-MUMs), but may contain mismatches according to the palindromic seed patterns. As with
multi-MUMs, any portion of a unique LMA may be non-unique.
We refer to the complete set of local multiple alignments generated in this step as
$\mathbf{M}$.

\subsubsection*{Guide tree construction}
\vspace{-0.2cm}
We compute a genome-content distance matrix and Neighbor Joining phylogeny
using the method described in~\citep{Henz2005},
adapted to work on local multiple alignments instead of BLAST hits.  We midpoint-root
the resulting tree to yield our progressive alignment guide tree $\Psi$.  Steps 2 and 3 in Figure~\ref{fig:algorithm} illustrate guide tree construction.

\subsubsection*{Local alignment anchor scoring}
\vspace{-0.2cm}
For a given local multiple alignment $m$ which includes genomes $g_i$ and $g_j$, we
compute a pairwise substitution score using a substitution matrix, which defaults to
the HOXD matrix~\citep{Chiaromonte2002}.
The HOXD matrix appears to discriminate well
between homologous and unrelated sequence in a variety of
organisms, even at high levels of sequence divergence.

The substitution matrix score quantifies the log-odds ratio that a pair of nucleotides share common ancestry, but
does not account for the inherent repetitive nature of genomic sequence.
Our desire to discriminate between alignment anchors that suggest orthology (or
xenology) and alignments of regions with random similarity or
paralogy requires that we somehow consider repetitive genomic sequence in our anchoring
score~\citep{Lippert2004}.  To do so, we compute a quantity termed $S_{occ}(g_i, x)$,
which is the number of occurrences in $g_i$ of the spaced seed pattern at site $x$.

We then combine the traditional substitution score for a pair of nucleotides with
an adjustment for the uniqueness of the aligned positions:

\begin{equation}\label{eqn:pairwiseColumnScore}
s(g_i, x, g_j, y) = \left\{ \begin{array}{ll}Q(g_i, x, g_j, y)&\mathrm{if}~Q(g_i, x, g_j, y) \leq 0 \\
&\\
\frac{2Q(g_i, x, g_j, y)}{S_{occ}(g_i, x) \cdot S_{occ}(g_j, y)} - Q(g_i, x, g_j, y) & \mathrm{otherwise} \end{array}\right.
\end{equation}

where $Q(g_i, x, g_j, y)$ is defined as the substitution matrix score for the
nucleotides at sites $x$ of genome $g_i$ and $y$ of $g_j$.   The product of
uniqueness scores for the two aligned positions approximates the number of possible
ways that repetitive sites could be combined.  For example, consider a repeat element
present in both genomes with copy number $r_i$ in genome $g_i$ and copy number $r_j$
in $g_j$.  There are $r_i r_j$ possible pairs of repeats.  When a pair of
nucleotides in a repeat element have a positive substitution score, the product
$S_{occ}(g_i, x) \cdot S_{occ}(g_j, y)$ down-weights the score.

We refer to the total pairwise anchor score for match $m$ in genomes $g_i$ and $g_j$ as $S(m, g_i, g_j)$; simply defined as the sum of scores for each alignment column:
\begin{equation}\label{eqn:localAnchorScore}
S(m, g_i, g_j) = \sum_{c = 1}^{m.Length}s(~g_i, pos(m, g_i, c), g_j, pos(m, g_j, c)~)
\end{equation}

where $pos(m,g_i,c)$ is defined as the position in $g_i$ represented by column $c$ of local alignment $m$.
In summary, the local alignment scoring scheme assigns high scores to well-conserved regions that are
unique in each genome and does not consider gap penalties.

\subsubsection*{Locally collinear blocks}
\vspace{-0.2cm}
A pair of genomes $g_i$ and $g_j$ may have undergone numerous genomic rearrangements
since their most recent common ancestor.  As such, local alignments among orthologous segments
of $g_i$ and $g_j$ may align segments that occur in a different order or orientation in each genome.
We define a pairwise Locally Collinear Block (LCB) as a subset of local alignments in $\mathbf{M}$ that occur in the same order and orientation in a pair of genomes $g_i$ and $g_j$, i.e. they are free from internal rearrangement.  To define pairwise LCBs among genomes $g_i$ and $g_j$, we first define the projection of the current set of Local Multiple Alignments $\mathbf{M}$ onto $g_i$ and $g_j$ as $\mathbf{M}:\overbrace{g_i,g_j}$, computed by setting all
left-end coordinates for genomes $\{g_i,g_j\}\smallsetminus\mathbf{G}$ to 0.  In alignment $m$,
for example, we would compute the projection onto $g_i$ and $g_j$ by setting all values in the left-end coordinate tuple $\langle x_1, x_2, \dots, x_G\rangle$ to 0 except for $x_i$ and $x_j$.  We can then minimally
partition the pairwise alignment projections $\mathbf{M}:\overbrace{g_i,g_j}$ into locally collinear blocks using the well-known breakpoint analysis procedure~\citep{gril, breakpointPhylogenies}.
We refer to the minimal partitioning of $\mathbf{M}:\overbrace{g_i,g_j}$ into locally collinear blocks as
$Lcb(\mathbf{M}:\overbrace{g_i,g_j})$, and the resulting subsets of $\mathbf{M}:\overbrace{g_i,g_j}$ as $L_{i,j}^1\cdots L_{i,j}^n \in Lcb(\mathbf{M}:\overbrace{g_i,g_j})$.

\begin{figure}
\centering
\includegraphics{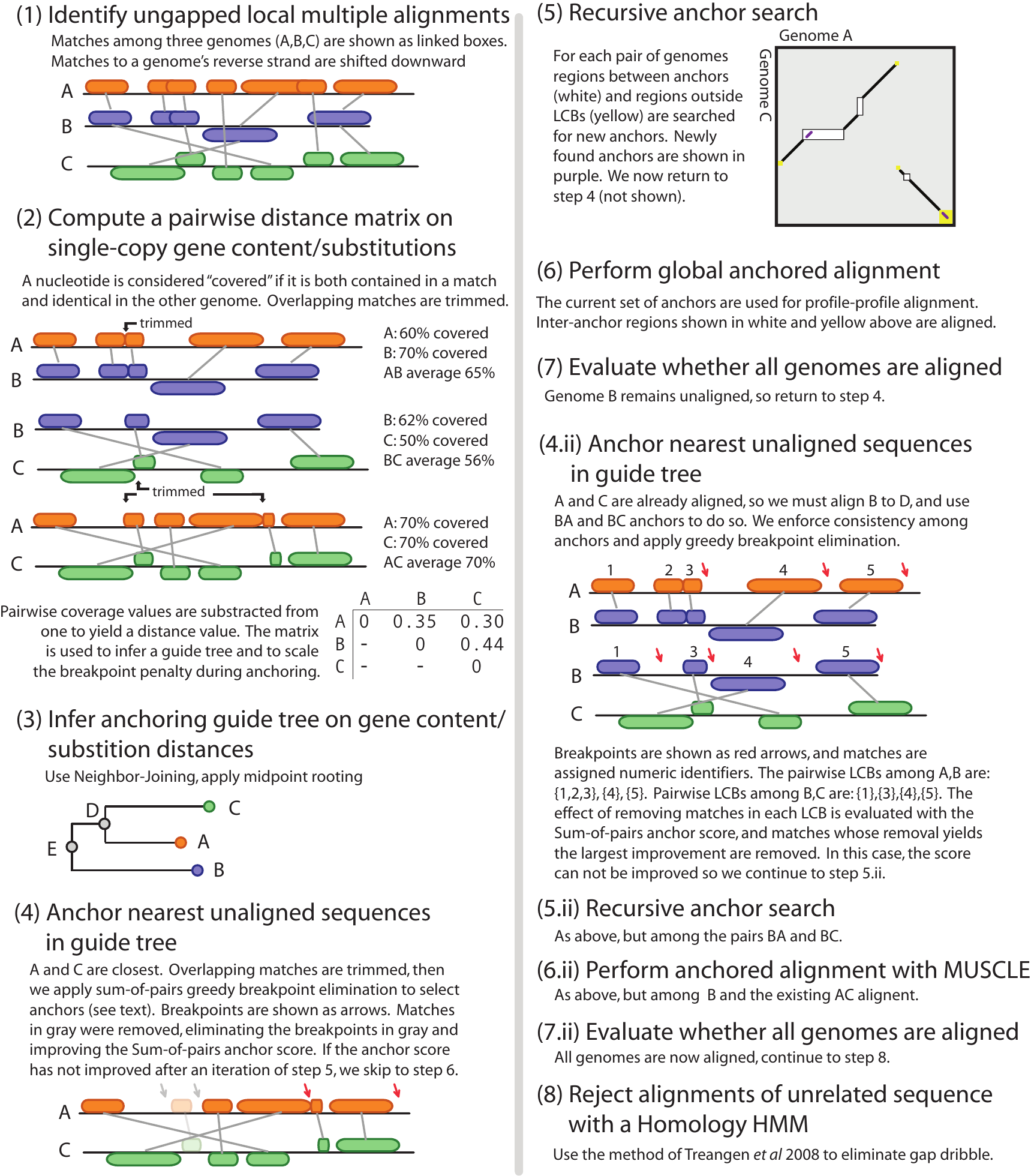}
\vspace{-0.2cm}
\caption{Overview of the alignment algorithm using three example genomes A, B, and C.} \label{fig:algorithm}
\end{figure}

\subsubsection*{Sum-of-pairs anchoring objective function}
\vspace{-0.2cm}
Our method computes anchored alignments progressively according to a guide tree.  When
computing an alignment at a given internal node of the guide tree $\Psi_n$,
we optimize the following sum-of-pairs LCB anchoring objective function
to select a set of alignment anchors:
\begin{equation}\label{SpAnchorScore}
SPAnchorScore(\Psi_n, \mathbf{M}) = \sum_{g_i \in
\mathcal{L}(\Psi_n.c_1)}\sum_{g_j \in \mathcal{L}(\Psi_n.c_2)}
\left\{-b|Lcb(\mathbf{M}:\overbrace{g_i,g_j})| +
\sum_{m\in \mathbf{M}:\overbrace{g_i,g_j}} S(m, g_i, g_j)
 \right\}
\end{equation}
where $|Lcb(\mathbf{M}:\overbrace{g_i,g_j})|$ refers to the total number of LCBs in the set of local alignments
among genomes $g_i$ and $g_j$.  The coefficient $b$ is a breakpoint penalty, which when multiplied by
the number of LCBs $|Lcb(\mathbf{M}:\overbrace{g_i,g_j})|$, creates a scoring penalty that increases in magnitude when the anchors in $\mathbf{M}:\overbrace{g_i,g_j}$ induce a larger number of LCBs.  The value of the breakpoint penalty $b$ is a user-controlled parameter in our implementation of the algorithm, and we use a default value of 30,000 as manual experimentation on real genome sequence data suggests this value represents a good tradeoff between sensitivity to small genomic rearrangements and filtration of spurious alignments.  Thus, anchor sets inducing fewer breakpoints are given higher scores.

\subsubsection*{Optimizing the SP anchoring objective function: greedy breakpoint elimination}
\vspace{-0.2cm}
We apply a greedy breakpoint elimination heuristic to optimize $SPAnchorScore(\Psi_n, \mathbf{M})$
which removes potential anchors from $\mathbf{M}$ until the score can no longer be increased.
Specifically, if all local alignments comprising an LCB $L_{i,j}$ are removed from $\mathbf{M}$, then
the number of LCBs in $Lcb(\mathbf{M}_{i,j})$ will decrease by at least one and at most four if
neighboring LCBs coalesce~\citep{mauve}.  The decreased number of LCBs, and hence breakpoints,
reduces the total breakpoint penalty in $SPAnchorScore(\Psi_n, \mathbf{M})$.

Our algorithm repeatedly identifies the LCB whose removal from $\mathbf{M}$ would provide the largest increase to $SPAnchorScore(\Psi_n, \mathbf{M})$ and removes the local alignments which comprise that LCB from $\mathbf{M}$.  This procedure corresponds to step 4 in Figure~\ref{fig:algorithm}.  Formally, we identify $L_{i,j}$ which satisfies:
\begin{equation}\label{greedyMove}
    \max_{g_i \in \mathcal{L}(\Psi_n.c_1),~g_j \in \mathcal{L}(\Psi_n.c_2)}
    \left\{ \max_{L_{i,j} \in Lcb(\mathbf{M}:\overbrace{g_i,g_j})}
    SPAnchorScore(\Psi_n, \mathbf{M}\smallsetminus L_{i,j})\right\}
\end{equation}
and remove the local alignments in $L_{i,j}$ from $\mathbf{M}$.  The greedy breakpoint elimination
process repeats until further removal of LCBs fails to improve the SP anchoring score.

\subsubsection*{Recursive anchoring}
\vspace{-0.2cm}
The initial set of local alignments in $\mathbf{M}$ is typically computed using
a seed weight that finds local alignments in unique regions of high sequence identity.
As such, the initial set of anchors frequently misses
homologous regions with lower sequence identity.  After anchor selection by greedy breakpoint
elimination (Equation~\ref{greedyMove}), our method searches for additional local alignments
between anchors existing among all pairs of genomes in $\mathcal{L}(\Psi_n.c_1)$ and $\mathcal{L}(\Psi_n.c_2)$,
see Figure~\ref{fig:algorithm} step 5.
To improve sensitivity during recursive anchor search, smaller seed weights
are used as described by~\citet{mauve}. Any new local alignments are added to $\mathbf{M}$.
After the recursive anchor search, we apply greedy breakpoint elimination to optimize the SP anchor score
once again. The recursive anchoring process repeats until $SPAnchorScore(\Psi_n, \mathbf{M})$ no longer improves.

\subsubsection*{Anchored profile alignment and iterative refinement}
\vspace{-0.2cm}

Each LCB at $\Psi_n$ contains alignment anchors, which we use to perform an anchored profile-profile
global alignment using modified MUSCLE software~\citep{muscle}.  After the initial profile-profile alignment,
we apply window-based iterative refinement to improve the alignment.  Step 6 of Figure~\ref{fig:algorithm} corresponds to this process.
Importantly, MUSCLE attempts to refine the alignment on a multitude of alternative guide trees and is not
restricted to the guide tree chosen for progressive anchoring.  The use of multiple guide trees is a particularly important feature in microbial genomes, which are subject to extensive lateral gene transfer.  It should be noted that our use
of MUSCLE as a refinement step is an approach used in other software pipelines as well~\citep{Margulies2007}.

\subsubsection*{Rejecting alignments of unrelated sequences}
\vspace{-0.2cm}
Segments of DNA between high-scoring alignment anchors are often unrelated, especially in bacteria.
Despite that, our method (like most other genome aligners)
applies a global alignment algorithm to all inter-anchor segments, na\"{\i}vely assuming that homology
exists.  Our assumption of homology often proves erroneous, so to arrive at an accurate alignment
we must detect forced alignment of unrelated sequence.  To do so, we apply an HMM posterior decoder
that classifies columns in a pairwise alignment as either homologous or unrelated.  The HMM structure,
transition, and emission probabilities are described elsewhere~\citep{Treangen2008}.  The HMM makes predictions of pairwise homology, which we combine using transitive homology relationships.
Regions found to be unrelated are removed from the final alignment.  Application of the
homology HMM is the final step in the alignment procedure, shown as step 8 in Figure~\ref{fig:algorithm}.

\subsubsection*{Implementation}
\vspace{-0.2cm}
The alignment algorithm has been implemented in the \texttt{progressiveMauve} program
included with Mauve v2.0 and later.  The program is open source C++ code (GPL), with 32- and 64-bit
binaries for Windows, Linux, and Mac OS X available from \url{http://gel.ahabs.wisc.edu/mauve}.
Time-consuming portions of the algorithm have been parallelized using OpenMP 2.5~\citep{openmp}.
An accessory visualization program is included (see Figure~\ref{fig:intergenic_viz}).
Default alignment parameters have been calibrated for bacterial genomes~\citep{Darling2006}.

\section*{RESULTS}
\subsection*{Quantifying alignment accuracy}
\vspace{-0.2cm}
Our new alignment algorithm uses approximations and computational heuristics
to compute alignments.  To understand the quality of alignments produced by our approach
it is essential to objectively quantify alignment accuracy.
Without a known `correct' genome alignment, automated alignment heuristics
can not be evaluated for accuracy.  Although several benchmark data sets exist for
protein sequence alignment~\citep{balibase,muscle}, no such
benchmark data sets exist for genome alignment with rearrangement.
Thus far, manual curation of a whole-genome multiple alignment that includes
rearrangement and lateral gene transfer has proven too time-consuming
and difficult.  Despite the lack of a manually curated correct alignment,
we can estimate the alignment accuracy by modeling evolution and aligning simulated data sets.  All results described in this section and the programs used to generate them are available as supplementary material.

\subsubsection*{Simulated evolution model}
\vspace{-0.2cm}
In previous work, we constructed a genome evolution simulator that
captures the major types, patterns, and frequencies of
mutation events in the genomes of Enterobacteriacae~\citep{mauve}.  We use the same
simulated model of evolution in the present study but with different evolutionary parameters.~
Given a rooted phylogenetic tree and an ancestral sequence we
generate evolved sequences for each internal and leaf node of the
tree, along with a multiple sequence alignment of regions conserved
throughout the simulated evolution. Along the branches, mutations
such as nucleotide substitution, indels, gene gain/loss, and inversion rearrangements~
are modeled as a marked Poisson process.    We score calculated
alignments against the correct alignments generated during the evolution process.

Although gene duplication does occur in bacteria, we do not \textit{explicitly} model it
here as duplications tend to be unstable in bacterial chromosomes and are
often counterselected~\citep{Achaz2003}.
Instead, we indirectly account for gene duplications in two ways.
First, the source DNA sequence for gene gain events comes from a 1Mbp pool of sequence.
At moderate to high simulated rates of gene gain, many megabases of DNA are sampled from
the donor pool, and as a result, identical donor sequence gets inserted into the simulated
genomes in multiple places.  The effect is similar to a dispersed repeat family, such as bacterial
IS elements or mammalian SINE elements.

Second, we use the genome sequence of \textit{E. coli} O157:H7 as ancestral
sequence and as donor sequence for all insertion and gene gain events.
The \textit{E. coli} O157:H7 genome has numerous naturally occurring repeats that
are carried on to simulated descendant genomes.  By using real genome sequence as
ancestral sequence, the resulting evolved genomes often have similar nucleotide,
dinucleotide, $k$-mer composition, repeat copy number and repeat distribution.  The unknown natural
forces governing the evolution of such traits
would otherwise be extremely difficult to capture in
a simulation environment.

Our experimental results at high mutation rates should be interpreted with caution, however, since the more
simulated mutations applied, the less a simulated genome
will look like a real genome.

\subsubsection*{Accuracy evaluation metrics}
\vspace{-0.2cm}
Previous studies of alignment accuracy have used a sum-of-pairs scoring scheme to
characterize the residue level accuracy of the
aligner~\citep{balibase,mauve}.  The experiments presented here use
sum-of-pairs scoring, but we also define new accuracy
measures to quantify each alignment system's ability to
predict indels and breakpoints of genomic rearrangement.
For each type of mutation, we define True Positive (TP), False Positive (FP), and False Negative (FN)
predictions as discussed below.  Using these definitions, we can measure the aligner's Sensitivity as $\frac{\mathrm{TP}}{\mathrm{TP}+\mathrm{FN}}$ and Positive Predictive Value (PPV) as $\frac{\mathrm{TP}}{\mathrm{TP}+\mathrm{FP}}$.

For nucleotide pairs, a TP is a
pair aligned in both the calculated and correct alignments.  A
FP is a nucleotide pair in the calculated alignment that is absent from
the correct alignment.  Likewise, a FN is a pair in the correct alignment
not present in the calculated alignment.  We do not quantify True Negative
(TN) alignments  as the number of TN possibilities is extremely large, growing with the product of sequence lengths.

\begin{figure}
\centering
\includegraphics[width=6in]{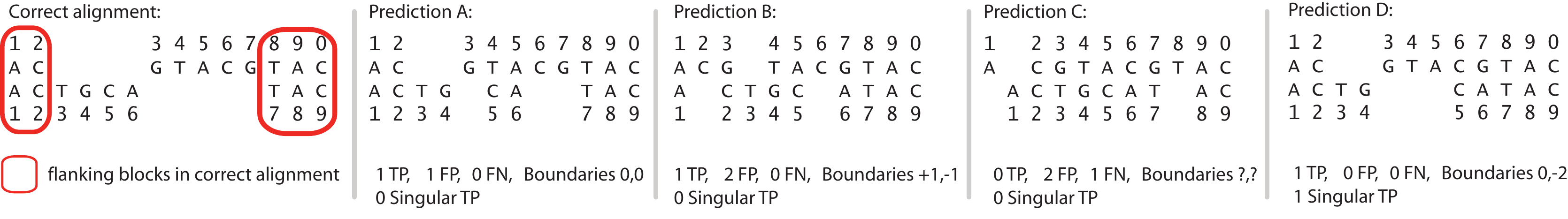}
\caption{Quantifying indel accuracy.  The correct alignment is shown at
left and four possible predicted alignments are shown as A, B, C, and D.  Nucleotides have been assigned a numerical identifier.  The
correct alignment has a single indel which partitions the alignment into three
sections: the left aligned block, the indel, and the right aligned block.  Predicted
alignments must have one correctly aligned nucleotide pair in each of the
three sections to count a true positive indel prediction.} \label{fig:indel_accuracy}
\end{figure}

We classify each indel in the correct alignment as a TP or a FN based on the predicted alignment.
A true positive indel has at least one correctly aligned nucleotide pair in the diagonal/block on
either side of the indel and at least one nucleotide correctly aligned to a gap within the indel (see Figure~\ref{fig:indel_accuracy}).
The number of TP indels will never exceed the number of indels in the correct alignment.
We define FP indel predictions as the number of excess indel predictions beyond the true positives.
FN indels lack a correctly predicted nucleotide pair in the flanking diagonals/blocks or
lack predictions of gaps in the correct gapped region.
Figure~\ref{fig:indel_accuracy} gives examples of each case.

Aligners are notoriously bad at predicting the exact position of indels~\citep{Lunter2008}.  Under our definition, a TP indel prediction need not predict the exact boundaries of an indel,
merely the existence of an indel.  This scheme allows us to distinguish cases of missing indel predictions from cases where the indel was predicted but not positioned correctly. We quantify indel boundary prediction accuracy as the
distance between the true boundary and the nearest aligned nucleotide pair in the diagonal/blocks
which flank the predicted indel.  When the predicted indel is too large, our metric assigns a positive value
to the boundary score.  When the predicted indel is too small, a negative value is assigned.

Large indels have historically caused problems for nucleotide aligners, which have a tendency
to break up large indels into a string of smaller gaps with intermittent aligned sequence.  Under our definition, a large indel can still be considered as a TP prediction even if it is broken into a string of smaller gaps by the aligner (See Figure~\ref{fig:indel_accuracy} prediction A for an example).  Our rationale is that the aligner did correctly predict the presence of unrelated sequence, for which it garners a TP, but erroneously predicts additional transitions to and from homology, which are classified as FP indel predictions.  To distinguish whether a TP indel was broken into two or more smaller gaps, we define a class of ``singular'' TP indel predictions as indels that were predicted as a single alignment gap.  See Figure~\ref{fig:indel_accuracy} prediction D for an example of a ``singular'' TP indel.

For each pair of genomes we also measure whether the aligner correctly predicts LCBs
among that pair, yielding a sum-of-pairs LCB accuracy metric.
For each pairwise LCB in the true alignment, we record a TP LCB prediction
when the predicted alignment contains at least one correctly aligned
nucleotide pair in that LCB.  Pairwise LCBs lacking any
correctly predicted nucleotide pairs are FN predictions. Finally,
pairwise LCBs in the predicted alignment lacking any correctly
aligned nucleotide pairs are False Positive (FP). Again, we do not measure TN.

As with indels, we define a separate metric to quantify how well each aligner
localizes the exact breakpoints of rearrangement.  For TP LCB predictions,
we record the difference (in nucleotides) between the boundaries of the
correct LCB and those of the predicted LCB.   The resulting value is negative
when the predicted LCB fails to include the full region of homology, and
positive when a predicted LCB extends beyond the true boundary.

Under our definitions of TP, FP, TN, and FN predictions, specificity, which is commonly defined as $\frac{\mathrm{TN}}{\mathrm{FP}+\mathrm{TN}}$, is not a useful metric.  The extremely large values taken on by TN would drive the quotient to 1 in most cases.

\subsubsection*{Accuracy on collinear genomes}
\vspace{-0.2cm}
\begin{figure}
\centering
\includegraphics[width=6.5in]{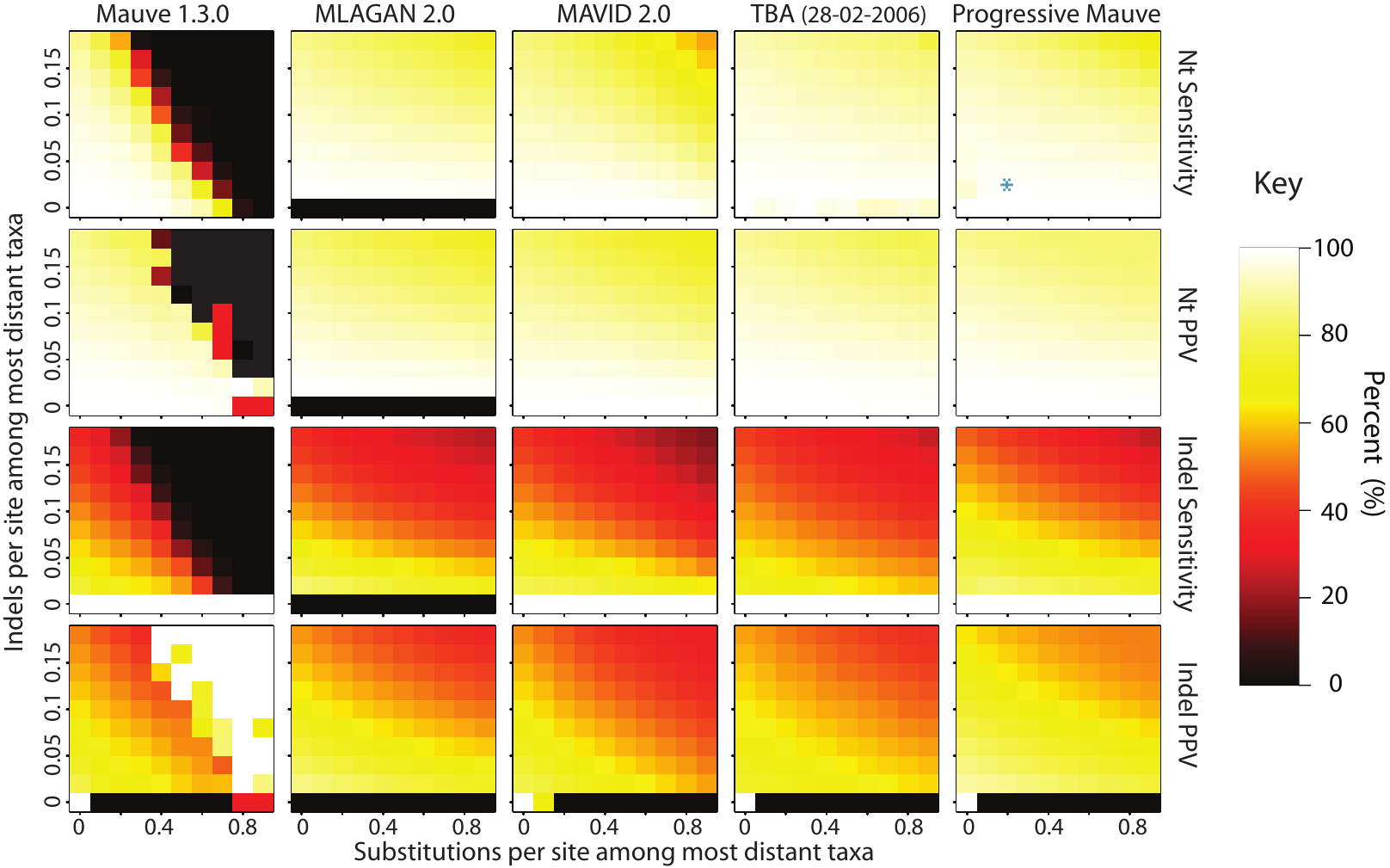}
\vspace{-0.2cm}
\caption{The accuracy of aligners on sequences evolved without rearrangement and with increasing nucleotide
substitution and indel rates. Aligners were tested on 100 combinations of indel
and substitution rate, with performance averaged over three replicates.  All methods lose accuracy as mutation rates grow, and the most accurate alignment method depends
on the particular mutation rates.  Progressive Mauve and MLAGAN exhibit the best indel sensitivity and positive predictive value (PPV), while TBA is more sensitive than other methods at extremely high mutation rates.  MLAGAN did not align genomes without indels within the allotted 10 hours,
resulting in the black row at the bottom.
The asterisk in this figure indicates the combination of indel rate and substitution rate
expected to be similar to our 23 target genomes.}\label{fig:ntsub_indel}
\end{figure}

Our first experiment compares the accuracy of Mauve 1.3.0, Progressive Mauve, MLAGAN 2.0, MAVID 2.0, and TBA 28-02-2006
when aligning collinear sequences that have undergone
increasing amounts of nucleotide substitution and indels.  For each combination
of indel and substitution rate, nine genomes are evolved from a 1Mbp ancestor
according to a previously inferred phylogeny~\citep{mauve} shown in Figure~\ref{fig:ntsub_inv_flux}.
We then construct alignments of evolved sequences using each aligner with default parameters, and quantify
sensitivity and positive predictive value, (PPV)
for nucleotide pair and indel predictions.  Three replicates were performed, results shown in
Figure~\ref{fig:ntsub_indel}.

In general, all aligners perform well on collinear sequence, except for Mauve 1.3.0 which is unable to anchor genomes with high mutation rates.  Of the tested aligners, TBA offers the highest nucleotide sensitivity, and Progressive Mauve gives the best indel sensitivity and positive predictive value in most cases.  Despite that, all aligners are quite bad at predicting indels accurately, which may be in part due to an inherent loss-of-information introduced during the course of simulated evolution~\citep{Lunter2008}.  We did not test the Pecan aligner here, although a detailed evaluation of its performance can be found elsewhere~\citep{Margulies2007} and we do perform some testing on it below.

\subsubsection*{Accuracy in the face of rearrangement and gene flux}
\vspace{-0.2cm}
\begin{figure}
\centering
\includegraphics[width=6.5in]{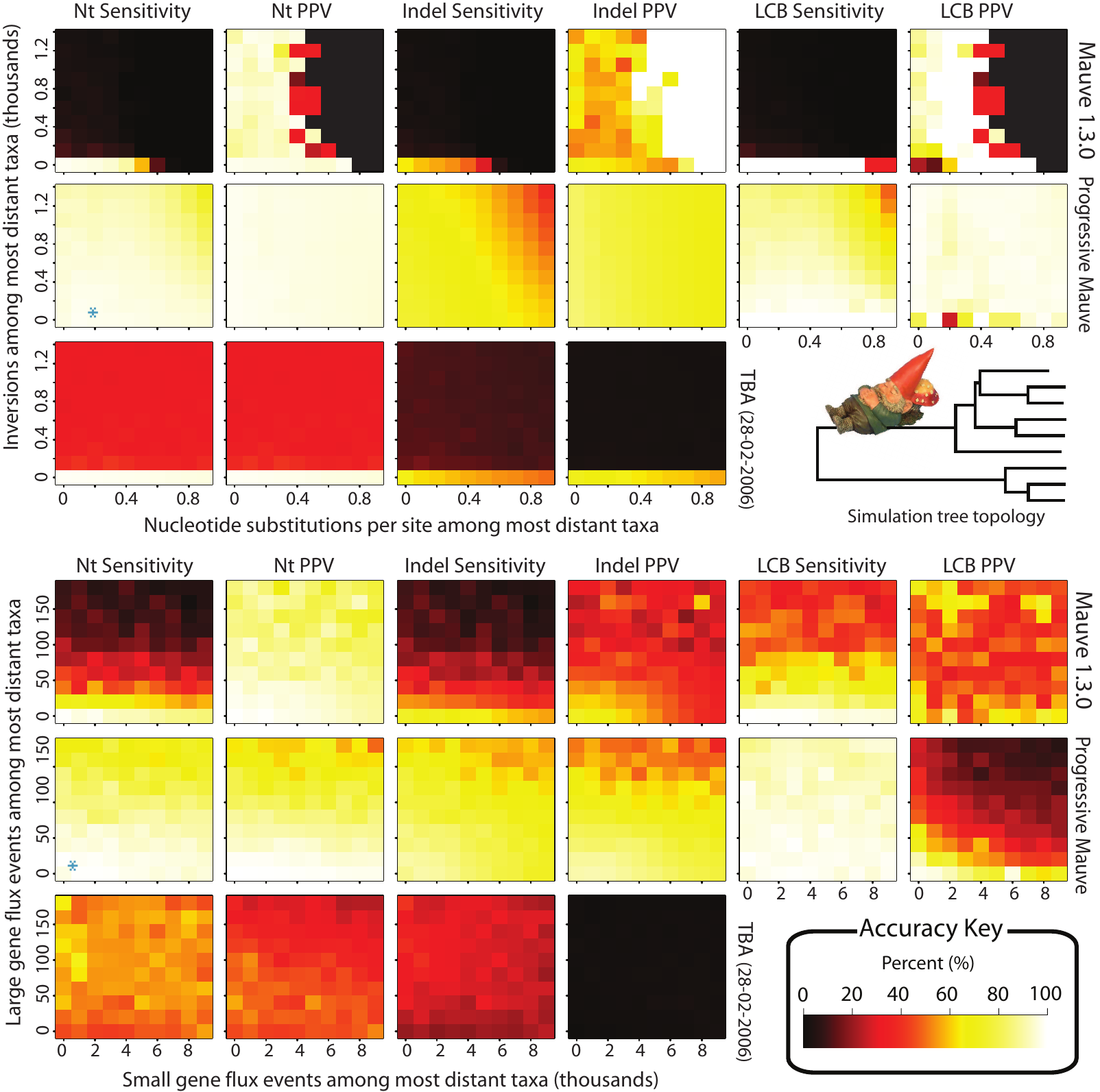}
\vspace{-0.2cm}
\caption{Accuracy of Mauve, Progressive Mauve, and TBA
when aligning genomes with inversions and gene flux. In the
experiments shown at top, the inversion rate increases along the
$y$-axis and the substitution
rate along the $x$-axis.  The most distant taxa have 0.05 indels per site.
Progressive Mauve
clearly outperforms Mauve 1.3.0 and TBA over the entire space of
inversion rates.  It should be noted that in the UCSC browser alignments
TBA is used in conjuction with a separate synteny-mapping method to identify rearrangements~\citep{Miller2007}, so the performance results given here are not cause for alarm.
Experiments at bottom quantify aligner performance in the presence
of small- and large-scale gene flux (gain/loss) events.
The $y$-axis gives the average number of
large gene flux events [length$\sim$Unif(10kbp, 50kbp)] between the most distant taxa,
while the $x$-axis gives small gene flux events [length$\sim$Geo(200bp)].
Substitution and indel rates are those indicated by the asterisk
in Figure~\ref{fig:ntsub_indel}, and the most distant taxa have 42 inversions on
average.  The asterisk in this figure indicates a simulation scenario
expected to be similar to our 23 target genomes.
Once again Progressive Mauve outperforms other methods, but all methods break
down when faced with substantial large-scale gene flux.  Of note, when Mauve 1.3.0
attains high PPV it usually does so with very poor sensitivity.}
\label{fig:ntsub_inv_flux}
\end{figure}
We assessed the relative performance of Mauve 1.3.0, Progressive
Mauve, and TBA~\citep{tba} when aligning genomes with high rates of genomic rearrangement,
gene flux, and nucleotide substitution.  Although the original TBA manuscript did
not fully describe alignment with genomic rearrangement, the most recent release (dated 28-02-2006)
handles it~\citep{aba,mulan}.  For our first set of experiments, shown in the top half of
Figure~\ref{fig:ntsub_inv_flux}, we simulated evolution at 100 combinations of substitution and
inversion rate.  In addition to nucleotide and indel accuracy, we also quantify LCB
accuracy on this data set.  The results indicate that Progressive Mauve
can accurately align genomes with substantially higher rates of
rearrangement than other approaches.  Although TBA exhibits lackluster performance in comparison
to Progressive Mauve, comparison with the results for MAVID 2.0 and MLAGAN 2.0 (shown in Supplementary Figure~\ref{fig:mlagan_mavid_inv_flux}) demonstrates that for all rates of inversion, TBA produces much better alignments than methods which assume genomes are free from rearrangement.

For the second set of experiments we simulated genomes with 10 increasing rates of small-scale gene flux and
10 increasing rates of large-scale gene flux.  Small gene flux events are
geometrically size distributed with mean 200bp, while large gene flux events have uniform lengths
between 10kbp and 50kbp.  These sizes were chosen to match empirically derived estimates~\citep{mauve}.
The results, shown in Figure~\ref{fig:ntsub_inv_flux},
indicate that Mauve 1.3.0 falters when faced with large-scale
gene flux, while Progressive Mauve and TBA perform significantly better.
As gene flux rates increase in our model, the amount of
orthologous sequence shared among genomes deteriorates, eventually
reaching zero in the limit of infinitely high gene flux rates.

\subsubsection*{\textit{Gap dribble} and the quality of long gap predictions}
Gene gain and loss events manifest themselves in genome alignments as long gaps.  Every predicted alignment gap implies at least one insertion or deletion of nucleotides has taken place in the history of the organisms under study.  Since we would like to quantify the contribution of gene flux to the target genomes, it is imperative that predicted alignment gaps be as accurate as possible.

Current sequence alignment methods typically score pairwise alignments with an affine gap scoring scheme consisting of a gap open penalty and a gap extend penalty.  In a probabilistic setting, the optimal affine-gap alignment corresponds to a viterbi path alignment from a pair-HMM with a single pair of insert and delete states~\citep{Durbin1998}.  However, when aligning genomes which have undergone a significant amount of gene gain and loss, an excess of large gaps exists that does not fit the gap size distribution imposed by a standard global alignment pair-HMM~\citep{Lunter2007}.  The net result is that under the affine gap model, aligners tend to break up large gaps into a series of small gaps interspersed with short stretches of improperly aligned nucleotides.  In the spirit of classifying systematic alignment errors introduced by \cite{Lunter2008}, we refer to this problem as \textit{gap dribble}, since short alignments are dribbled along the large gap.  The large number of small gaps creates problems when trying to reconstruct the history of gene gain and loss events, since they imply a much greater number of insertions and deletions than actually occurred.

\begin{figure}
\centering
\includegraphics[width=6.5in]{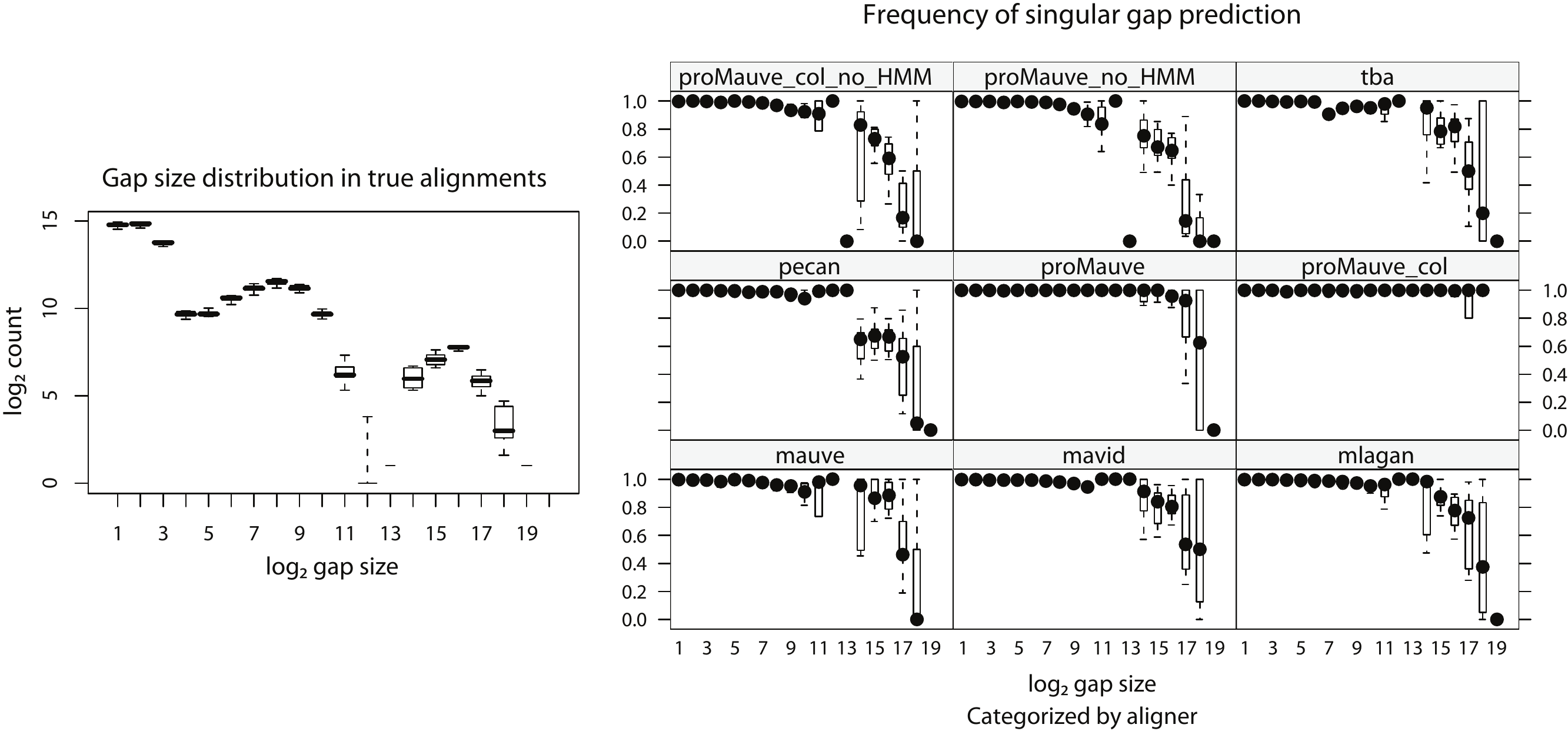}
\vspace{-0.2cm}
\caption{\textbf{Left} Average size distribution of gaps in an alignment of the nine genomes evolved
at mutation rates which correspond to previous estimates for the
\textit{E. coli}, \textit{Shigella}, and \textit{Salmonella}.  The gap size distribution was averaged over
10 simulations. \textbf{Right} Fraction of TP indel predictions that are singular TP indel predictions by true gap size.  Ten replicate simulations of evolution with gene gain, gene loss, indels, and nucleotide substitution were performed and alignments were computed using each aligner.  Predicted indels were classified according to the definitions given in Figure~\ref{fig:indel_accuracy}, namely, a singular True Positive implies the true gap is predicted as a single gap.  Remaining True Positive indels have the true gap broken up into two or more predicted gaps.  For each aligner, the fraction of singular predicted gaps is shown as a function of gap size.  All aligners do well in predicting small gaps, but large gaps present problems.  Most aligners, including Pecan which uses an extra pair-HMM state to model long gaps, tend to predict long gaps as a series of short gaps interspersed with alignments of unrelated sequence.  We refer to such behavior as ``gap dribble.'' Progressive Mauve was run with default parameters (proMauve), without the Homology HMM (proMauve\_no\_HMM), with the option to assume genomes are collinear (proMauve\_col), and finally assuming collinearity and without the HMM (proMauve\_col\_no\_HMM). }
\label{fig:mind_the_gaps}
\end{figure}

Using our simulated evolution platform, we quantify the performance of each aligner in predicting gaps of varying size.  We simulated evolution of collinear genomes (no rearrangement) that have undergone a realistic amount of gene gain and loss, corresponding to previous estimates for the rates of these events in the enterobacteria~\cite{mauve}.  Nucleotide substitutions and indels were modeled to occur at the rate indicated by the blue asterisk in Figure~\ref{fig:ntsub_indel}, and gene gain and loss events were modeled to occur with twice the frequency indicated by the blue asterisk in Figure~\ref{fig:ntsub_inv_flux}.  Figure~\ref{fig:mind_the_gaps}~Left gives the observed size distribution of gaps.

We then applied each aligner to the simulated genomes and measured the accuracy of gap predictions as a function of gap size.  The aligners Mauve 1.3.0, MAVID 2.0, MLAGAN 2.0, TBA 28-02-2006, Progressive Mauve, and Pecan v0.7 were tested.  Pecan v0.7 is a new aligner that has been demonstrated to have excellent performance~\citep{Margulies2007,Paten2008} by virtue of using probabilistic consistency during the anchoring process.  Moreover, Pecan v0.7 uses a pair-HMM with an extra gap state specifically designed to model long indels.  The reconstructed alignments were scored against the true alignments and results for ten replicates were recorded.

The right side of Figure~\ref{fig:mind_the_gaps} shows the quality of each aligner's indel predictions as a function of the true gap size.  Shown is the frequency with which gaps of a particular size are predicted as a single gap (singular TP) instead of a string of smaller gaps with interspersed alignments of non-homologous sequence (nonsingular TP).  From the figure, it is obvious that aligners which use an affine gap penalty tend to perform poorly in predicting large gaps.  Somewhat surprisingly, the pair-HMM with an extra gap state used by Pecan to model long indels still yields poor predictions of long gaps, although sensitivity is quite good (not shown).  Progressive Mauve appears to perform well at all gap sizes, especially when the aligner is told to explicitly assume the genomes are collinear (proMauve\_col).  To determine whether Progressive Mauve's performance results from its anchoring algorithm or use of the Homology HMM to reject alignments of unrelated sequence,
we also tested Progressive Mauve without the Homology HMM, shown as panels proMauve\_no\_HMM and proMauve\_col\_no\_HMM.  Without the Homology HMM Progressive Mauve yields inferior results, indicating that the Homology HMM does indeed address the problem of gap dribble.

\section*{DISCUSSION}
\vspace{-0.2cm}

Progressive Mauve alignments enable a wide variety of downstream research.  Here we illustrate some applications with an alignment of 23 complete \textit{E. coli}, \textit{Shigella} and \textit{Salmonella} genomes.  The alignment can be used to characterize the shared (core) and total (pan-genome) amount of sequence found in these species.  In related work, we demonstrate a method to statistically infer the phylogenetic history of gene flux and identify sudden changes in the overall rate of gene flux in ancestral lineages (see Didelot \textit{et al} joint submission with this manuscript). The alignment can also be used to extract variable sites for more traditional phylogenetic analyses.  Finally, we demonstrate that Progressive Mauve identifies both conserved regulatory regions and hypervariable intergenic regions.

When applied to twenty-three \textit{E. coli}, \textit{Shigella}, and \textit{Salmonella} genomes, the resulting alignment reveals a core genome of 2,675 segments conserved among all taxa, which account for an average of 2.46 Mbp of each genome.  Between the core segments lie regions conserved among subsets of taxa and regions unique to individual genomes. By counting each core, unique, and subset segment exactly once, one constructs a total pan-genome that includes genes and intergenic regions alike. The 23 genomes have a pan-genome of 15.2 Mbp, approximately three times that of a single strain, indicating a tremendous degree of variability in both genic and intergenic content.

\begin{figure}
\centering
\centering
\includegraphics[width=6.5in]{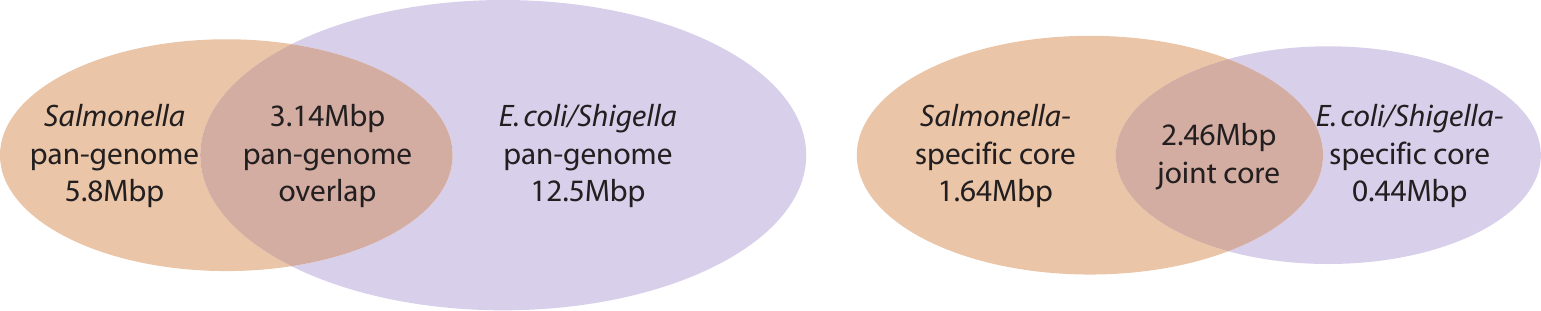}
\vspace{-0.4cm}
\caption{Venn diagram of the pan-genome (left) and core genome (right) of \textit{E. coli/Shigella} and \textit{S. enterica}.}\label{fig:venn_pancore}
\end{figure}

\textit{Shigella} spp. are widely recognized as \textit{E. coli} based phylogenetic analyses~\citep{Pupo2000} and genome comparisons~\citep{Yang2007}, though the original phenotypically derived taxonomy persists. We will refer to them collectively as \textit{E. coli/Shigella}.  Similarly, taxonomic revisions of \textit{Salmonella}, have collapsed almost all strains into a single species: \textit{S. enterica}.  Thus, we are examining the structure of the pan and core genomes of two sister species, \textit{E. coli/Shigella} and \textit{Salmonella}. The 16 \textit{E. coli/Shigella} strains have a pan-genome of 12.5 Mbp and core of 2.9 Mbp, while the seven \textit{S. enterica} serovars have a pan-genome of 5.8 Mbp and a core genome of 4.1 Mbp. The intersection of the core genomes is the joint core, while the union of the pan-genomes is the combined pan-genome, shown in Figure~\ref{fig:venn_pancore}.  Note that the intersection of pan-genomes is 580 kb larger than the joint core.  This counter-intuitive situation arises when components of the core-genome of one group are found in some, but not all members of the other species.  In this instance, 220 kb can be attributed to losses of genes in \textit{Shigella} strains that are otherwise conserved among all \textit{E. coli} and \textit{Salmonella}.  A more detailed dissection of the patterns of gene flux in the \textit{E. coli/Shigella} lineage appears in the joint Didelot \textit{et al} submission.

\subsubsection*{Inference of genome rearrangement history}
\vspace{-0.2cm}
Progressive Mauve alignments also make an excellent starting point for analysis of genome rearrangement patterns.  Genome rearrangement is known to occur via a multitude of mutational forces, including inversion, transposition, and duplication/loss, and is especially prominent in bacterial pathogens.  Methods already exist to infer inversion histories among pairs of genomes~\citep{Hannenhalli1995,Tannier2004} and multiple genomes~\citep{Tang2003,Larget2002}.  More general models to account for multiple chromosomes and multi-break rearrangements have also been developed~\citep{Yancopoulos2005,Bergeron2006,Alekseyev2008}, although not yet in either phylogenetic or Bayesian contexts.

Most genome rearrangement history inference methods do not also infer gene gain and loss, but instead assume that gene content across genomes is equal.  When gene content is nearly equal, current models can use a multiple genome alignment to infer patterns of selection on genome rearrangement~\citep{Darling2008}.  However, equal gene content has proved to be the exception rather than the rule.  Despite that, a Progressive Mauve alignment with differential content can be trivially reduced to contain only segments conserved among all taxa of interest, yielding a signed gene-order permutation matrix that is suitable for current genome rearrangement inference software.

\subsubsection*{Alignment visualization}
\vspace{-0.2cm}
Genome alignments are large and complex entities that are not usually suitable for direct interpretation.  Genome comparison browsers such as the UCSC browser~\citep{Miller2007}, VISTA~\citep{Dubchak2000}, and others have proven invaluable as tools to facilitate understanding of whole genome alignments. To aid in use of Progressive Mauve alignments, we have developed an interactive visualization program that can present a complex alignment in a meaningful and easily understandable visual paradigm.

\begin{figure}[t]
\centering
\includegraphics[width=6.5in]{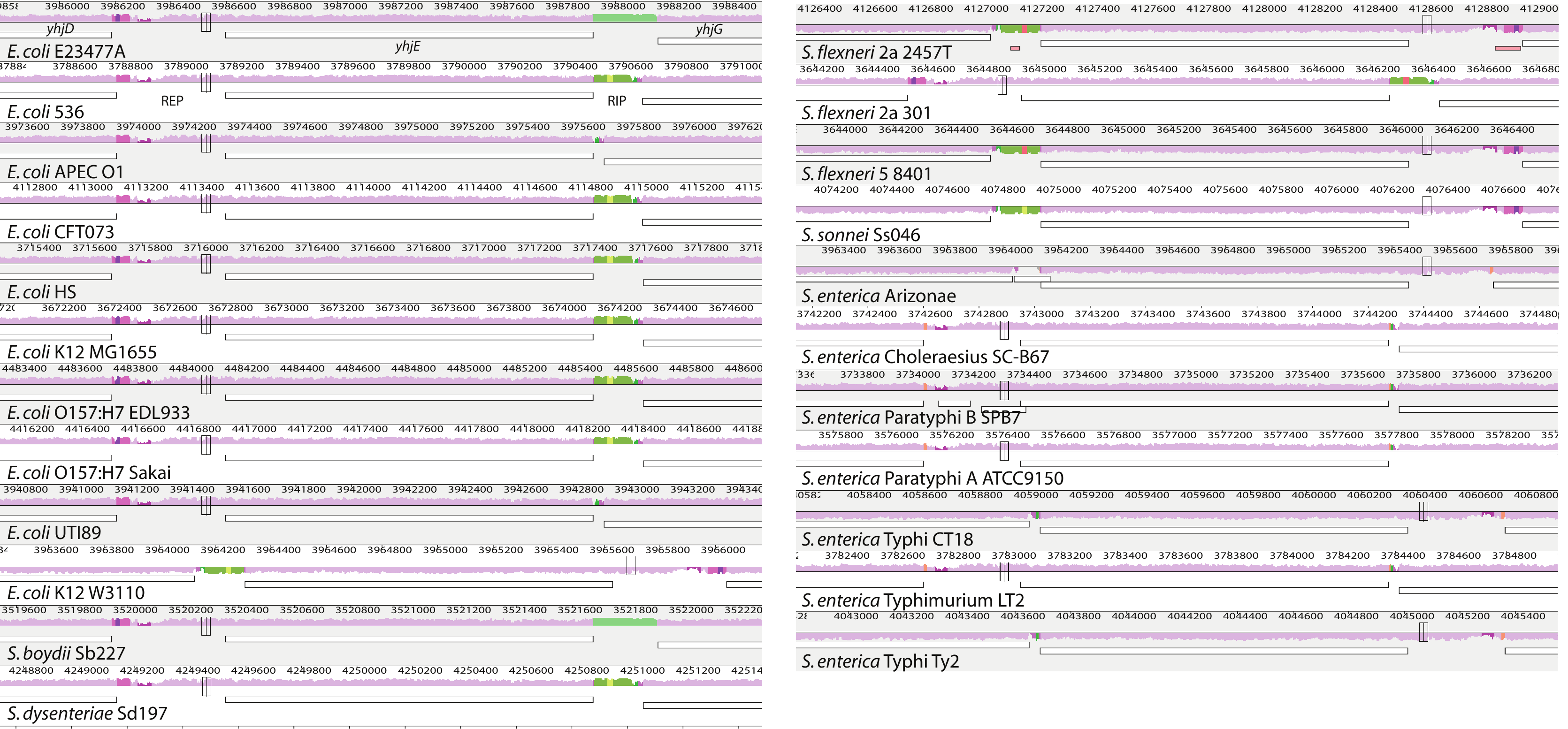}
\vspace{-0.2cm}
\caption{A Mauve visualization of the hypervariable intergenic regions surrounding \textit{yhdE}. Each genome is laid out in a horizontal track,
with annotated coding regions shown as white boxes.  A colored similarity plot is shown for each genome, the height of which is proportional to the level of sequence identity in that region.  When the similarity plot points downward it indicates an alignment to the reverse strand of the genome.  Colors in the similarity plot indicate the combination of organisms containing a particular segment of the genome.  Segments colored pink/mauve are conserved among all organisms, while purple segments are conserved in everything but \textit{Salmonella}, and segments colored in olive green are conserved among non-uropathogenic \textit{E. coli}.  The visualization system is interactive and written in Java, and works on all computers supporting Java 1.4 or later.}
\label{fig:intergenic_viz}
\end{figure}

The visualization system illustrates three major aspects of genome evolution: genome rearrangement, patterns of segmental gain and loss, and selective pressures for or against regional conservation of nucleotide sequences.  Figure~\ref{fig:intergenic_viz} illustrates two of these aspects in a visualization of the 23-way alignment of \textit{E. coli}, \textit{Shigella}, and \textit{Salmonella}.

Figure~\ref{fig:intergenic_viz} shows the region surrounding the \textit{yhjE} gene, which encodes a product in the Major Facilitator Superfamily of transporters.  \textit{yhjE} is flanked by \textit{yhjD} to the left, and \textit{yhjG} to the right.  The intergenic regions between these three genes are hypervariable (as indicated by the variety of colors in Figure~\ref{fig:intergenic_viz}) and have been subject to multiple insertion and deletion events.  The hypervariable nature of the regions surrounding \textit{yhjE} may not be surprising, because it harbors a REP element to the left, and a RIP element to the right.  REP elements contain a series of two or more 35-bp palindromic repeats and are known for a variety of functions, including the binding of DNA gyrase and PolI, and as mRNA anti-decay hairpins or rho-dependent attenuators.  RIP elements are a specialized form of REP elements that contain an IHF binding site, and this particular RIP also contains a REPt transcription terminator sequence.  IHF is a global transcriptional regulator in \textit{E. coli}.

Interestingly, the patterns of insertion/deletion in the intergenic regions surrounding \textit{yhjE} do not follow the expected taxonomic patterns, suggesting instead that recombination among strains has taken place.  The RIP region present to the right of \textit{yhjE} in most \textit{E. coli} has been replaced with an unrelated sequence in \textit{E. coli} E23477A and \textit{S. boydii} (shown as turquoise in Figure~\ref{fig:intergenic_viz}), but not in \textit{S. sonnei}.  Those three strains form a clade in the \textit{E. coli}/\textit{Shigella} taxonomy~\citep{Didelot2008} with \textit{E. coli} E23477A branching first, so convergent evolution must have occurred here.

The pattern of intergenic variability surrounding \textit{yhjE} suggests potential regulatory divergence, a much studied evolutionary mechanism in eukaryotes largely overlooked in prokaryotic research.  The \textit{yhjE} locus is by no means the only region harboring intergenic variablity; a screen of the 23-way alignment identifies 102 other strictly intergenic regions with similarly variable conservation patterns.

\subsubsection*{Conclusions}
\vspace{-0.2cm}
We have presented a novel multiple genome alignment heuristic which demonstrates a substantial accuracy improvement on simulated datasets.  Key features of the approach are an anchor scoring function that penalizes alignment anchoring in repetitive regions of the genome and penalizes genomic rearrangement.  Use of a
Sum-of-pairs approach enables robust scoring of genomes that have undergone both gene flux and rearrangement---a scenario not addressed by previous alignment methods.

Future efforts to improve genome alignment may explicitly incorporate models of evolutionary distance into alignment scoring process~\citep{Fu2007}.  Multiple alignment methods based on probabilistic consistency have demonstrated great promise in the context of amino-acid alignment~\citep{probcons} and aligning collinear genomic regions~\citep{Paten2008}, and in principle, could be also extended to genome alignment with rearrangement.

No method reconstructs error-free genome alignments, and any particular alignment is likely to contain errors that can substantially influence downstream inference.  However, methods to estimate the confidence in aligned columns are under continuing development~\citep{Lunter2008}.  Downstream inference methods that can explicitly cope with the inherent uncertainty in reconstructed alignments will be crucial for continued advances in comparative genomics.

\subsection*{ACKNOWLEDGEMENTS}
\vspace{-0.2cm}
We thank Eric Cabot, Anna Rissman, and Paul Infield-Harm for suggestions and contributions related to alignment visualization.
Work supported in part by NIH R01-GM62994 to N.T.P.
and NSF grant DBI-0630765 to A.E.D.  This project has also been
funded in part with Federal funds from the National Institute of
Allergy and Infectious Diseases, National Institutes of Health,
Department of Health and Human Services, under Contract No. HHSN266200400040C.
\bibliographystyle{apalike}
\bibliography{proMauve_gr}

\clearpage

\section*{APPENDIX}
\begin{table}[b]
  \centering
  \small
\begin{tabular}{|r|c|c|}
\hline Organism & Genome size w/Plasmids & Accession \\
\hline\hline
\textit{E. coli} K-12 MG1655 & 4,654,221 & U00096 \\
\hline
\textit{E. coli} K-12 W3110 & 4,646,332 & AP009048 \\
\hline
\textit{E. coli} HS & 4,643,538 & AAJY00000000 \\
\hline
\textit{E. coli} O157:H7 EDL933 & 5,623,806 & AE005174 \\
\hline
\textit{E. coli} O157:H7 Sakai & 5,594,477 & BA000007 \\
\hline
\textit{E. coli} E24377A & 4,980,187 & AAJZ00000000 \\
\hline
\textit{E. coli} CFT073 (UPEC) & 5,231,428 & AE014075 \\
\hline
\textit{E. coli} UTI89 (UPEC) & 5,179,971 & CP000243 \\
\hline
\textit{E. coli} APEC O1 & 5,082,025 & CP000468 \\
\hline
\textit{E. coli} 536 (UPEC) & 4,938,920 & CP000247 \\
\hline
\textit{Shigella boydii} Sb227  & 4,646,520 & CP000036 \\
\hline
\textit{Shigella flexneri} 2a 2457T  & 4,988,914 & AE014073  \\
\hline
\textit{Shigella flexneri} 2a 301  & 4,828,821 & AE005674 \\
\hline
\textit{Shigella flexneri} 5 8401  & 4,574,284 & CP000266 \\
\hline
\textit{Shigella dysenteriae} Sd197  & 4,551,958 & CP000034 \\
\hline
\textit{Shigella sonnei} Ss046  & 5,039,661 & CP000038 \\
\hline
\textit{Salmonella enterica} Choleraesius B67  & 4,944,000 & AE017220 \\
\hline
\textit{Salmonella enterica} Typhi Ty2  & 4,791,961 & AE014613 \\
\hline
\textit{Salmonella enterica} Typhi CT18  & 5,133,713 & AL513382 \\
\hline
\textit{Salmonella enterica} Typhimurium LT2  & 4,951,371 & AE006468 \\
\hline
\textit{Salmonella enterica} Paratyphi A ATCC9150  & 4,585,229 & CP000026 \\
\hline
\textit{Salmonella enterica} Arizonae  & 4,600,800 & NC\_010067 \\
\hline
\textit{Salmonella enterica} Paratyphi B SPB7 & 4,858,887 & NC\_010102 \\
\hline
\end{tabular}
\vspace{0.2cm}
  \caption{\textbf{(Supplementary)}Twenty-three publicly-available, finished genome sequences from the
  genera \textit{Salmonella}, \textit{Escherichia}, and \textit{Shigella} form our target  set for multiple genome alignment.}
\label{table:supp_enterics}
\end{table}

\begin{figure}[t]
\centering
\includegraphics{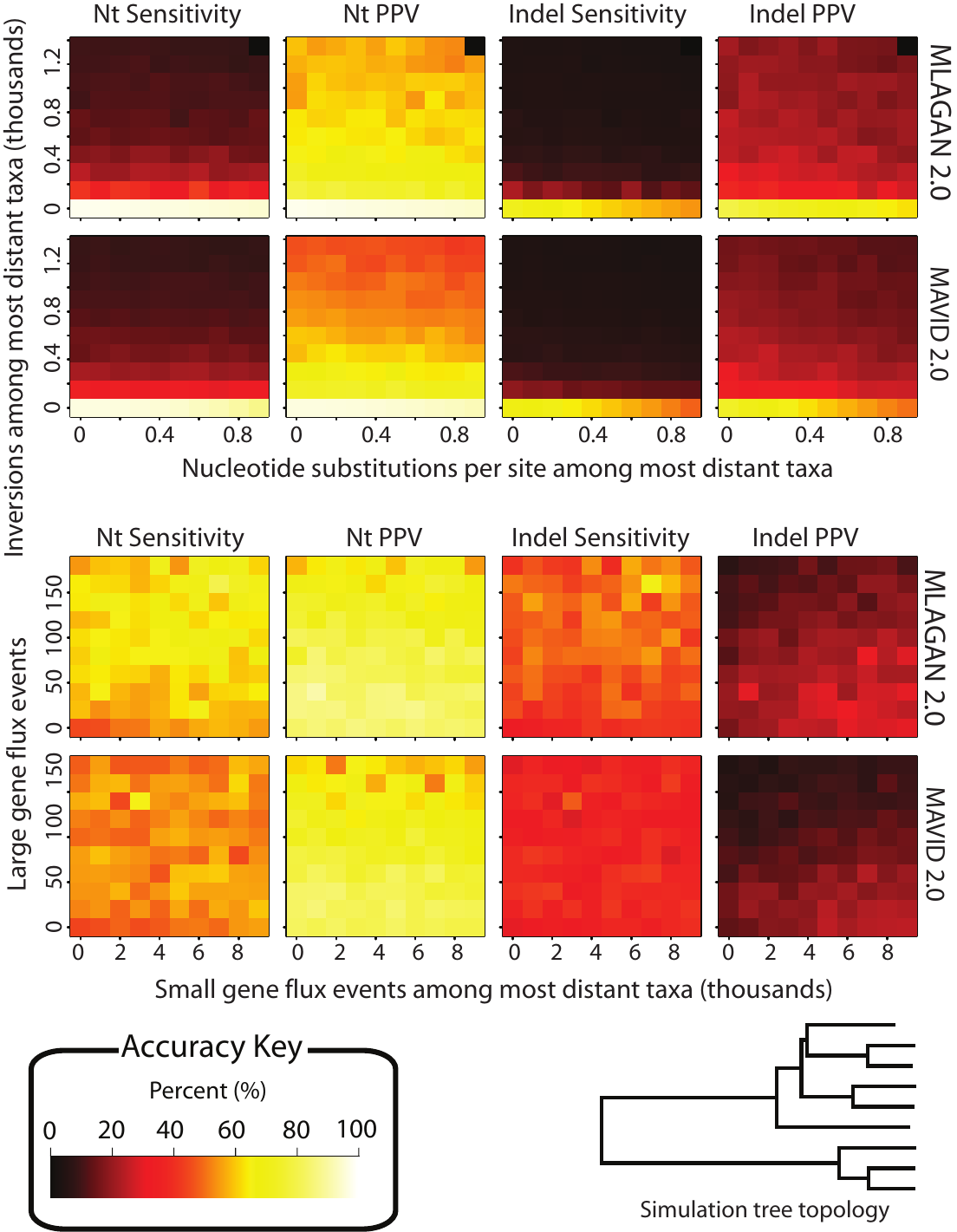}
\caption{\textbf{Supplementary}~The performance of MLAGAN 2.0 and MAVID 2.0 on genomes evolved with inversions, gene gain, and gene loss corresponding to Figure~\ref{fig:ntsub_inv_flux}.  Neither MLAGAN nor MAVID were designed to align genomes with rearrangement and differential content; this figure demonstrates the decay in alignment quality when such forces are present in the data but not modeled by the aligner.  The black dot in the upper right corner of the MLAGAN results indicates that MLAGAN did not finish alignments at the highest combination of inversion and substitution rate within the allotted 10 hour limit.
}
\label{fig:mlagan_mavid_inv_flux}
\end{figure}

\end{document}